\def\beq{\begin{equation}} 
\def\eeq{\end{equation}} 
\def\bed{\begin{displaymath}} 
\def\eed{\end{displaymath}} 
\def\beqq{\begin{eqnarray}} 
\def\eeqq{\end{eqnarray}} 
\def\bedd{\begin{eqnarray*}} 
\def\eedd{\end{eqnarray*}}

\def\n{\nonumber}
 
\def\bbbz{{\bf Z}} 
 
\def\bbb1{{\rm 1\!1}}

\newcommand{\refs}[1]{(\ref{#1})}

\documentstyle[12pt]{article} 
 
\advance\hoffset by -.5in 
\begin{document} 
\bibliographystyle{unsrt} 
 
\def\pr{\prime} 
\def\pa{\partial} 
\def\es{\!=\!} 
\def\ha{{1\over 2}} 
\def\>{\rangle} 
\def\<{\langle} 
\def\mtx#1{\quad\hbox{{#1}}\quad} 
\def\lam{\lambda} 
\def\La{\Lambda} 
 
\def\A{{\cal A}} 
\def\dal{\dot\alpha}
\def\de{\delta}
\def\be{\beta}
\def\dbe{\dot\beta}
\def\ga{\gamma}
\def\G{\Gamma} 
\def\Ga{\Gamma} 
\def\F{{\cal F}} 
\def\J{{\cal J}} 
\def\M{{\cal M}} 
\def\R{{\cal R}} 
\def\W{{\cal W}} 
\def\tr{\hbox{tr}} 
\def\al{\alpha} 
\def\d{\hbox{d}} 
\def\De{\Delta} 
\def\L{{\cal L}} 
\def\H{{\cal H}} 
\def\Tr{\hbox{Tr}} 
\def\I{\hbox{Im}} 
\def\e{\epsilon}
\def\R{\hbox{Re}} 
\def\h{\bar h}
\def\di{\partial\!\!\!\!\,/}
\def\ti{\int\d^2\theta} 
\def\bti{\int\d^2\bar\theta} 
\def\ttbi{\int\d^2\theta\d^2\bar\theta} 
 \def\bD{\bar D}
\def\bh{\bar h}
\def\la{\lambda}
\def\dla{\dot\lambda}
\def\dde{\dot\delta}
\def\bp{\bar\phi}
\def\W{{\cal{W}}}
\def\cL{{\cal{L}}}

\begin{center}

{\Large\bf On Metric Perturbations in Brane-World 
Scenarios}\\
\vspace*{2cm}
Andrey Neronov and Ivo Sachs \\
\vspace*{1cm}
{\em Theoretische Physik\\
Ludwig-Maximilians Universit\"at\\
Theresienstrasse 37 \\80333 Munich\\ Germany}\\
\vspace*{2cm}
\end{center}
\abstract{In this note 
we reconsider linearised metric perturbations in the 
one-brane Randall-Sundrum Model. We present a simple 
formalism to describe metric perturbations caused by matter 
perturbations on the brane 
and remedy some misconceptions concerning the constraints imposed on 
the metric and matter perturbations by 
the presence of the brane.}

\vspace*{2cm}


An interesting alternative to standard Kaluza-Klein compactification is 
to view our $4$-dimensional Universe as a $3$-brane embedded in a bigger 
space with large, or infinite extra dimensions, but such that matter fields are 
localised on the brane \cite{Rubakov1}-\cite{Visser}. 
Considerable effort has been devoted to investigate possible 
observable consequences of such scenarios. 
For this one has to determine how the gravitational dynamics, 
including matter is affected by 
the extra dimensions in which gravity can propagate. 
The linearised metric perturbation in Brane-World Scenarios have been 
considered by numerous authors 
\cite{RS1,RS2}, \cite{Garriga}-\cite{0010215}.
 
In \cite{Garriga,GKR} the authors considered linear 
perturbations of a Minkowski brane in coordinates in which the 
four-dimensional metric is transverse-traceless. In this gauge 
the brane appears ``bent'' in the presence of matter, that is, 
the coordinates are discontinuous on the brane. In \cite{arefeva} and 
\cite{K} the same analysis was carried out in alternative gauges 
for which the coordinates are continuous across the 
brane. On the other hand linear perturbations were analysed for an 
expanding 
Universe on the brane 
\cite{9912233}-\cite{0010215} and some 
cosmological consequences were analysed using a generalisation of the 
$3+1$ 
decomposition of the metric perturbations \cite{Bardeen,Kodama,Mukhanov}. In 
this note we present a simple generalisation of the formalism 
\cite{K} which allows us to relate and correct some of the results 
obtained 
within the different approaches for metric perturbations caused by matter on 
a 
Minkowski brane. Concretely we find the general solution for linear metric 
perturbations generated by arbitrary matter perturbations on the brane in 
the axial gauge $H_{\mu 4}\es 0$. There is a degeneracy 
in the system of linear perturbation equations due to the remaining scalar 
scalar gauge-freedom. Exploring the space of solution for the linear 
perturbations we may identify various 
previous solutions found in different gauges. On the other hand 
we find that the so-called generalised longitudinal gauge 
for the scalar perturbations \cite{0005032} can not be imposed for generic 
matter perturbations on the brane. We find the approach presented here 
very well suited for perturbations induced by matter on the brane, however,  
we do not address primordial perturbations in this paper. Also we leave 
the 
generalisation to expanding brane cosmologies for future work.

To describe the metric perturbations consider the Ansatz 
\beqq
ds^2=g_{AB}dx^A dx^B&=&e^{2A(z)}\left(\eta_{AB}+H_{AB}\right) dx^A dx^B \ ,\n\\
A&=&-\log(1+\kappa|z|)\ ,
\eeqq
which, in the absence of matter satisfies the background Einstein 
equations\footnote{In units with $8\pi G_5\es 1$; $A\es 0,\cdots,4$; 
$\mu\es 0,\cdots,3$.} 
\beq
{\cal{G_{AB}}}= \Lambda g_{AB}+V\eta_{\mu\nu}\de(z)\ ,
\eeq
provided 
\beq
\Lambda=6\kappa^2\mtx{and} V=-6\kappa\ .
\eeq
Next we consider the linearised Einstein equations for the perturbation 
$H_{AB}$ (in the gauge $H_{\mu4}\es 0$) 
in the presence of matter localised on the brane. 
After a series of standard 
manipulations and Fourier transformation along the brane 
we end up with the system of  equations 
\beqq\label{linmom}
\ha p^2 H_{\mu\nu}-\ha H''_{\mu\nu}+\ha p_\mu p_\nu H_{44}- p_\mu 
p_{(\nu}H_{\mu)\lambda}+&&\n\\
\frac{4\kappa^2}{(1+\kappa z)^2}\eta_{\mu\nu} H_{44}+
\frac{\kappa}{2(1+\kappa z)}\left(H'\eta_{\mu\nu}-H'_{44}
\eta_{\mu\nu}+3H'_{\mu\nu}\right)&=&0\n\\
-\ha p_\mu H'+\ha p^\lambda H'_{\lambda\mu}-p_\mu
\frac{3\kappa}{2(1+\kappa z)}H_{44}&=&0\\
\ha p^2 H_{44}-\ha H''-\frac{\kappa}{2(1+\kappa z)}
\left(H'_{44}-H'\right)+\frac{4\kappa^2}{(1+\kappa z)^2}H_{44}&=&0\ ,\n
\eeqq
for $z>0$, together with the jump conditions at $z\es 0$
\beq
-\ha [H'_{\mu\nu}]=\kappa H_{44}\eta_{\mu\nu}+T_{\mu\nu}-
\frac{1}{3}T\eta_{\mu\nu}\ .
\eeq
Here prime denotes the derivative alon $z$, $T_{\mu\nu}$ is the matter 
stress tensor and $p^2\es p^\mu p_\mu$. Following \cite{K} we substitute the general Ansatz for $H_{\mu\nu}$ and 
$H_{44}$ perturbations on the brane
\beqq\label{ansatz}
H_{\mu\nu}&=&a(p,z)T_{\mu\nu}+b(p,z)T\eta_{\mu\nu}+c(p,z)p_\mu p_\nu T\n\\
H_{44}&=&d(p,z)T\ ,
\eeqq
Note that \refs{ansatz} is not 
the general solution to \refs{linmom} as we can always add a solution of 
the homogeneous equations $H^0_{AB}$ to \refs{ansatz}. In what follows we 
will set $H^0_{AB} \es 0$, that is, we restrict ourself to perturbations 
generated by matter on the brane. Substitution of \refs{ansatz} into 
\refs{linmom} then leads to the set of equations \cite{K}
\beqq\label{2p}
&&a'+3b'+\frac{3\kappa}{1+\kappa z}d=0
\eeqq
\beq\label{28}
p^2\left(a+3b\right)-\frac{\kappa}{1+\kappa z}
\left(a'-3p^2c'\right)=0
\eeq
\beq\label{26}
a''-\frac{3\kappa}{1+\kappa z}a'-p^2a=0
\eeq
\beq\label{29}
a+2b-c''+\frac{3\kappa}{1+\kappa z}c'+d=0\ ,
\eeq
supplemented with the jump conditions 
\beq\label{jump}
\ha[a']=-1\mtx{;}\ha[b']=\frac{1}{3}-\kappa d\ ,
\eeq
and $c'$ is continuous across $z=0$. The 
$\bbbz_2$-symmetry $z\to -z$ implies that $c'|_{0}\es 0$. 

It turns out that the system of equations \refs{2p}-\refs{29} is degenerate. 
Here we relate this degeneracy with the remaining scalar gauge freedom 
by comparing the Ansatz \refs{ansatz} with the general 
Ansatz \cite{0005032,WAN}. 
\beq\label{gim}
g_{AB}=e^{2A(z)}\pmatrix{-(1+2\Phi)&-W_{0i}&-W\cr
-W_{0i}&(1-2\Psi)\de_{ij}+2p_ip_jE+2p_{(i}F_{j)}+h_{ij}&W_{4i}\cr
-W&W_{4i}&(1+2\Gamma)}\ ,
\eeq
where
\beq
W_{ai}=p_iB_a+S_{ai}\ , a=0,4\ ,
\eeq
and $S_{ai}$ is divergence free $p^i S_{ai}=0$. The diffeomorphisms 
$\xi^A$ are also decomposed into divergence free $3$-vectors $\xi^i$ 
($2$ degrees of freedom), and $3$ 
scalars\footnote{Note, however 
constraint  $\xi^5(y\es 0,x)\es 0$ in in the presence of a $3$-brane at 
$z\es 0$} 
$\xi,\xi^0,\xi^5$. In what follows we use the vector gauge degrees of 
freedom 
to set 
\beq
S_{4i}=0\ .
\eeq
Under the remaining diffeomorphisms the different scalars in 
\refs{gim} transform as  
\beqq\label{gt}
&&\de\Psi=-A'\xi_4\mtx{;}\de\Phi=p_0\xi_0-A'\xi_4\mtx{;}
\de \Gamma =\xi_4'+A'\xi_4\\
&&\de B_0=\xi_0-p_0\xi\mtx{;}\de B_4= -\xi_4+\xi'\mtx{;}
\de E=\xi\mtx{;} \de W= p_0\xi_4+\xi'_0\ .\n
\eeqq
In order to compare the two Ans\"atze we 
further fix two scalar gauge degrees of freedom by imposing 
\beq
B_4=W=0\ ,
\eeq
which can always be chosen for matter localised on the brane (see also 
\cite{K}). To continue we compare the parametrisations \refs{gim} with \refs{ansatz} 
leading to 
the identifications 
\beqq\label{ident}
-2\Phi&=&a\rho-b T+cp_0T\ ,\n\\
S_{0i}&=&a Pe_{i}\ ,\n\\
B_0&=&cp_0T\ ,\n\\
-2\Psi&=&a p+b T-\frac{1}{3}a\Delta \tilde\pi\ ,\n\\
2E&=&a\tilde\pi+cT\ ,\n\\
F_i&=&a\pi_i\ ,\n\\
h_{ij}&=&a\bar\pi_{ij}\ ,\n\\
2\Gamma&=&dT\ .
\eeqq
Here, $\rho$ and $P$ are the density- and pressure perturbations on the 
brane and  $e_i$ is a transverse unit vector. The quantities 
$\tilde\pi,\pi_i$ and $\bar\pi_{ij}$ denote the scalar-, vector- and 
tensor part of the anisotropic stress and $a,b,c,d$ are the solutions 
of the system of equations \refs{2p}-\refs{29}. 
To complete the comparison between the Ansatz \refs{ansatz} 
and the parametrisation \refs{gim} we now 
compare the jump conditions found with the two Ans\"atze. 
The Ansatz \refs{ansatz} together with the jump conditions \refs{jump} 
leads to 
\beqq
\Phi'|_{z=0}&=&\frac{1}{3}\rho+\frac{1}{2}p-\kappa\Gamma\n\\
\Psi'|_{z=0}&=&\frac{1}{6}\rho-\kappa\Gamma\n \\
S'_{0i}|_{z=0}&=& -pe_i\n\\
F'_i|_{z=0}&=&-\pi_i\\
h'_{ij}|_{z=0}&=&-\bar\pi_{ij}\n\\
E'|_{z=0}&=&-\ha\tilde\pi\n\\
B'_0|_{z=0}&=&0\ .\n
\eeqq
For $\tilde \pi =0$ these jump conditions agree with those found in 
\cite{0005032,WAN} respectively, thus establishing the equivalence with 
the Ansatz \refs{ansatz} modulo homogeneous solutions of the linearised 
Einstein equations \refs{2p}-\refs{29}. 

As explained above we still have one scalar gauge transformation available. 
This allows, for example, to bring the metric into the form $B_0=0$ 
while keeping $W=B_4=0$. This is achieved by
\beq
\xi=\ha cT\mtx{;}\xi_4=\xi'\mtx{;}\xi_0=-\ha cp_0 T\ .
\eeq
This is the solution presented in \cite{K}. 

Next we discuss what happens if we try to fix the remaining scalar gauge 
freedom by imposing the so-called generalised longitudinal gauge 
$E=B_0=B_4=0$ \cite{0005032}. We do this by taking as a starting point 
the gauge $c\es 0$. The longitudinal gauge is then obtained with 
\beq
\xi=-\ha a\tilde\pi \mtx{;}\xi_4=\xi'\mtx{;}\xi_0=-\ha a\tilde\pi\ .
\eeq
Note that in this gauge $W=-a'p_0\tilde\pi$ no longer 
vanishes. More importantly,  
\beq
\xi^4(z= 0)=-\ha a'\tilde\pi=\ha\tilde\pi\neq 0 \ ,
\eeq
that is, this gauge is only compatible with having the brane at $z\es 0$ 
if 
$\tilde\pi\es 0$. This is the conclusion reached in \cite{0005032}. However, 
as we can see from the above this conclusion is gauge dependent as it 
only holds in the longitudinal gauge chosen in \cite{0005032}\footnote{We 
have been informed that this problem concerning the longitudinal gauge will 
also be discussed in \cite{W2}.}. 

Let us now see how the metric perturbation in a gauge with  
$H_{44}\es 0$ \cite{RS2,Garriga,GKR} are obtained in our approach. 
It is clear that this gauge corresponds to $d\es 0$. From \refs{gt} we 
see that $d\es 0$ can be achieved by
\beq
\xi'_4-\frac{\kappa}{(1+\kappa z)}\xi_4=-\frac{d}{2}T(p)\ .
\eeq
Using \refs{2p} we integrate this equation as 
\beq
\xi_4=\al_0(p)(1+\kappa z)+\frac{(1+\kappa z)}{6\kappa}(a+3b)T(p)\ ,
\eeq
where $\al_0(p)$ is an integration constant to be determined later. 
Without restricting the generality we can assume $c\es 0$ before 
performing 
the gauge transformation leading to $d\es 0$. In that case we can use 
\refs{28} to integrate the constraint $\xi_4\es \xi'$ as
\beq
\xi=\al_1(p)+\frac{\al_0(p)}{2\kappa}(1+\kappa 
z)^2+\frac{1}{6p^2}a(p,z)T(p)\ .
\eeq
Here, the integration constant $\al_1(p)$ corresponds to the 
$z$-independent 
four dimensional gauge transformation which we can set to zero at present 
(see also \cite{Garriga}). The remaining integration constant 
$\al_0(p)$ is, in turn, determined by the requirement that $\xi_4(p,0)
\es 0$ \cite{K}. This then implies $\al_0(p)=\frac{1}{6 p^2}T(p)$. 
Finally 
we obtain the function $c(p,z)$ using 
\refs{ident} and \refs{gt}, i.e. 
\beq\label{dc}
\de c(p,z) p_0 T(p)=-\de B_0=2p_0\xi\ , 
\eeq
up to a $z$-independent integration constant which we set to zero. Thus, 
\beq\label{dc1}
\de c(p,z)=-\frac{1}{3p^2}
\left( \frac{(1+\kappa z)^2}{2\kappa}+a(p,z)\right)\ .
\eeq
Correspondingly, $\de b(p,z)$ is given by
\beq\label{db}
\de b(p,z)=\frac{\kappa}{(1+\kappa z)}\frac{1}{3p^2}
\left(1+\kappa z+a'\right)\ .
\eeq
The function $a(p,z)$ is gauge-invariant. Note, however, that in 
this gauge eqn. \refs{2p} implies that $a+3b$ is independent of $z$. 
On the other hand $a'$ is bounded \cite{K} and 
hence \refs{28} implies that $c'$ 
diverges quadratically as $z\to\infty$, as previously observed 
\cite{Garriga,GKR,K}. Here, the large $z$ divergences of the linear 
perturbation 
$H_{AB}$ in this gauge are directly read off from from \refs{2p}-\refs{29}. 
As explained in \cite{Garriga,GKR,K} in order to remedy this divergence we 
need to relax the condition that the brane be situated at $z\es 0$. 
This then allows us to impose the extra conditions 
$p^\nu H_{\mu\nu}\es H^\mu_\mu\es 0$, that 
is, the Randall-Sundrum gauge. Indeed, as is clear from \refs{ansatz} 
these two conditions correspond to 
\beq \label{cond}
b+\de b+\de c p^2=a+4(b+\de b)+\de cp^2=0\ ,
\eeq
where we have used again that $c\es 0$ before the gauge transformation. 
Substituting the expressions \refs{dc1} and \refs{db} into \refs{cond} it is 
then easy to see that \refs{cond} is fulfilled for $\al_0\es 0$. 
The displacement of the brane, that is, $\xi_4|_{z\es 0}$ is then given by 
\beq\label{dis}
\xi_4(p,0)=-\frac{1}{6p^2}T(p)\ ,
\eeq
in agreement with \cite{Garriga,GKR}. In this gauge, 
$c'(p,0)T(p)$ 
measures the displacement of the brane. On the other hand $b$ and $c$ are 
expressed algebraically in terms of $a$ and hence $b$ and $c$ are bounded 
functions of $z$. Note, however, that this gauge violates the 
consistency condition $c'(p,0)\es 0$. This inconsistency is resolved 
by noting that in this gauge the coordinates are discontinuous on the brane, 
so that the Einstein equations \refs{linmom} are modified. 

\vspace{.8cm}
 
{\bf Note added: } After submitting the original version of this note 
we became aware of the latest version of \cite{K} which also discusses 
gauge aspects in the system dicussed here. 
We would like to thank Z. Kakushadze for pointing this out to us. 

 \vspace{.8cm}
 
We would like to thank R. Abramo and A. Barvinski for many enlightening 
discussions and we also thank K. Malik, V. Mukhanov and 
S. Solodukhin for helpful comments. This work was supported 
by the DFG-SPP 1096 f\"ur Stringtheorie and DFG-SFB 375 f\"ur 
Astroteilchenphysik.

\end{document}